\documentclass[12pt,preprint]{aastex}

\usepackage{graphicx}

\def \apj{{ApJ}}
\def \aj{{AJ}}
\def \mnras{{MNRAS}}
\def \nat{{Nature}}
\def \aa{{A\&A}}

\shorttitle{ALMA observations of the Vela pulsar}
\shortauthors{Mignani et al.}

\begin{document}

\title{The first detection of a pulsar with ALMA\thanks{}}

\author{
R. P. Mignani\altaffilmark{1,2}, 
R. Paladino\altaffilmark{3},
B. Rudak\altaffilmark{4}, 
A. Zajczyk\altaffilmark{5}, 
A. Corongiu\altaffilmark{6}, 
A. de Luca\altaffilmark{1}, 
W. Hummel\altaffilmark{7}, 
A.  Possenti\altaffilmark{6}, 
U. Geppert\altaffilmark{2}, 
M. Burgay\altaffilmark{6}, 
G. Marconi\altaffilmark{8} }

\affil{\altaffilmark{1} INAF - Istituto di Astrofisica Spaziale e Fisica Cosmica Milano, via E. Bassini 15, 20133, Milano, Italy}
\affil{\altaffilmark{2} Janusz Gil Institute of Astronomy, University of Zielona G\'ora, Lubuska 2, 65-265, Zielona G\'ora, Poland}
\affil{\altaffilmark{3} INAF - Istituto di Radioastronomia and Italian ALMA Regional Centre, via P. Gobetti 101, 40129 Bologna, Italy }
\affil{\altaffilmark{4} Nicolaus Copernicus Astronomical Center, Polish Academy of Sciences, ul. Rabia\'nska 8, 87100, Toru\'n, Poland }
\affil{\altaffilmark{5} University of Iowa, Department of Physics and Astronomy, 203, Van Allen Hall, Iowa City, IA 52242, USA } 
\affil{\altaffilmark{6} INAF Ð Osservatorio Astronomico di Cagliari, Via della Scienza 5, I-09047 Selargius, Italy}
\affil{\altaffilmark{7} European Southern Observatory, Karl Schwarzschild-Str. 2, D-85748, Garching, Germany}
\affil{\altaffilmark{8} European Southern Observatory, Alonso de Cordova 3107 Vitacura, Santiago de Chile, Chile}

\begin{abstract}
Although there is a general consensus on the fact that pulsars' radio emission is coherent in nature, whereas the emission from the optical to high-energy $\gamma$-rays is due to incoherent processes, it has not been established yet at which wavelengths the transition occurs, a key information for all emission models of pulsar magnetospheres. Of course, to address this issue  covering the spectral region between the GHz radio frequencies and the mid-infrared (IR) is crucial. 
We used the Atacama Large Millimetre Array (ALMA) to observe the Vela pulsar (PSR\, B0833$-$45), one of the very few  observed in radio and from the mid-IR up to the very high-energy $\gamma$-rays. We detected Vela at frequencies of 97.5, 145, 233, and 343.5 GHz, which makes it the first pulsar ever detected with ALMA. Its energy density spectrum follows a power-law of spectral index $\alpha  =  -0.93 \pm 0.16$. This corresponds to very high brightness temperatures - from $10^{17}$ to $10^{15}$ K - suggesting that a coherent radiative process still contributes to the mm/sub-mm emission.  Therefore, this is the first indication of coherent emission  in pulsars extending to the sub-mm range. At the same time, we identified an extended structure, preliminarily detected in ground-based near-IR observations, at a distance of $\sim 1\farcs4$ from the pulsar, possibly interpreted as a counter-jet protruding from the pulsar.
  \end{abstract}


\section{Introduction}

The Vela pulsar (PSR\, B0833$-$45), in the Vela supernova remnant, is a relatively young isolated neutron star, with a characteristic age $\tau \approx$ 11000 years, estimated from the ratio between its spin down period $P_{\rm s}$ (89 ms) and its derivative $\dot{P}_{\rm s}$ ($12.5\times10^{-14}$s s$^{-1}$) as  $\tau = P_{\rm s}/(2 \dot{P}_{\rm s})$. According to the magnetic dipole model (e.g., Pacini 1968),
the timing parameters (Manchester et al.\ 2005) yield a surface magnetic field $B_{\rm s} = 3.38\times10^{12}$ G and a rotational energy loss rate $\dot{E} = 6.96\times10^{36}$ erg s$^{-1}$. The Vela pulsar was discovered in the radio band in 1968 (Large et al.\ 1968) and soon after it was detected both in the X (Harnden et al.\ 1973) and $\gamma$-rays (Thompson et al.\ 1975) and, later on, also in the optical band (Lasker 1976; Wallace et al.\ 1977).  
Vela was also detected in 
the near-ultraviolet (UV) with the {\em HST} (Romani et al.\ 2005), the near-infrared (IR) with the 
VLT  (Shibanov et al.\ 2003), and the mid-IR with {\em Spitzer} (Danilenko et al.\ 2011) becoming, aside from PSR\, B0531+21 in the Crab Nebula, the only radio pulsar detected all the way from the mid-IR up to  $\gamma$-ray energies above 50 GeV, thanks to {\em Fermi} observations (Leung et al.\ 2014). 

The Vela pulsar spectral energy distribution (SED) is quite complex. In the soft X-ray band (0.1--2 keV), the SED features a thermal component  ascribed to emission from the cooling neutron star surface (Manzali et al.\ 2007).  The rest of the SED is purely non-thermal and is characterised by the combination of different  power law (PL) spectra $F_{\nu} \propto \nu^{\alpha}$, where $F_{\nu}$ is the flux density per unit frequency $\nu$, with turnovers between different energy bands.
In particular, a spectral  turnover is observed in the radio band at frequencies above a few  GHz, pheraphs associated with the transition between coherent and incoherent emission processes, with the former responsible for the radio emission and the latter for the non-thermal emission from the near-IR up to the high-energy $\gamma$ rays.

Determining the frequency  range where such transition occurs provides key information for the radiation emission models in pulsar magnetospheres. Of course, the knowledge of  the pulsar spectrum in the spectral region between mm and micron wavelengths,  i.e. between the high-frequency radio band and the mid-IR,  is crucial to constrain the value of the transition frequency.  Being one of the three pulsars detected in the mid-IR, with the Crab (Temim et al.\ 2009) and, possibly,  Geminga (Danilenko et al.\ 2011), Vela is an obvious target for observations with the Atacama Large Millimetre Array (ALMA), the most sensitive telescope in the 
3.6--0.3 mm range.

This manuscript is divided as follows: observations and data analysis are described in Section 2, whereas the results are presented and discussed in Sections 3 and 4, respectively. Summary and conclusions then follow in Section 5.

\section{Observations and Data Reduction}

We observed the Vela pulsar with ALMA during Cycle 3, from March to July 2016,  in four different frequency bands: Band 3 (97.5 GHz), Band 4 (145 GHz), Band 6 (233 GHz), and Band 7 (343.5 GHz), corresponding to wavelengths of 3.07, 2.07,  1.28, and 0.87 mm, respectively. 
 Our ALMA observations were executed 
 with the 12-m array, with an average  of 38 antennas, in different configurations, with maximum baselines ranging from 495 m (Band 7) to 867 m (Band 3), allowing a spatial resolution ranging from $\sim 0\farcs6$ to $0\farcs8$.  
The total time spent on the target was $\sim$ 3.3 hours, which provided a $1 \sigma$ sensitivity  of $\sim$ 20 $\mu$Jy/beam in each of the selected bands.  Table 1 reports the details of each observation.
The 12-m array correlator was configured to allow observations in dual polarisation  in a 7.5 GHz wide band, divided in four
spectral windows (1.875 GHz each), with 128 channels,  $\sim$ 15 MHz wide. For each band, standard frequencies $\nu$ have been used, which are optimised to achieve the best atmospheric transmission.  The observed field of view decreases with the frequency and ranges from 63\farcs4 in Band 3 to 18\arcsec\ in Band 7.
Raw data were calibrated using the Common Astronomy Software Applications (CASA) software package (version 4.5.3) and the ALMA pipeline (Pipeline-Cycle3-R4-B).  
To set the flux density scale for the Band 4, 6, and 7 observations, we used flux measurements of the quasar J1107$-$4449. The flux density uncertainty on this calibrator reported in the ALMA Calibrator Source Catalogue\footnote{https://almascience.eso.org/sc/} is 10\%.
For the Band 3 observations, we fixed the flux density scale using  data from Callisto observations, and using the Butler-JPL-Horizons 2012 standard model.  We, conservatively, assume the uncertainty on the flux density to be 10\%.
For the astrometry calibration we used as a reference the radio position of the quasar J0811$-$4929
at
$\alpha=08^{\rm h}  11^{\rm m} 08\fs803$ ($\pm 0\farcs0058$); $\delta  = -49^\circ 29\arcsec 43\farcs509$ ($\pm 0\farcs0021$)
at about $6^{\circ}$ from our target, as reported in the ALMA Calibrator Source Catalogue.
We produced images from the calibrated data using the task {\tt clean} in CASA. The images were generated using "natural''  weighting, in order to maximise the sensitivity. 
For all images, we computed the pulsar position at the epochs of our ALMA observations using its radio coordinates as a reference (Dodson et al.\ 2003), $\alpha=08^{\rm h}  35^{\rm m} 20.61^{\rm s}$; $\delta  = -45^\circ 10\arcsec 34\farcs87$ (epoch 2000.0), and correcting them for its radio proper motion (Dodson et al.\ 2003), $\mu_{\alpha}=-49.6$ mas yr$^{-1}$ and  $\mu_{\delta}=+29.9$ mas yr$^{-1}$. The uncertainties on the radio proper motion and pulsar position are negligible and they do not affect the precision on the pulsar position determination on the ALMA images.  The astrometric accuracy in the different bands, taken into account the signal--to--noise (S/N) and the resolution, is 56, 48, 43, 123 mas, in Band 3, 4, 6, and 7, respectively. 

\begin{table*}[tbh]
\begin{center}
\footnotesize{

\begin{tabular}{ccccccc}
\hline \noalign {\smallskip}
Band & $\nu$ & Date & Time  & Beam & rms & flux density \\
 & GHz & 2016 & min & arcsec $\times$ arcsec & $\mu$Jy/beam & $\mu$Jy\\

\hline \noalign {\smallskip}
3& 97.5 & July 23& 22& 0.75 $\times$ 0.68 & 25 & 252$\pm$35\\
4& 145 & July 14 & 22 & 0.77 $\times$ 0.74 & 20 & 170$\pm$26\\ 
6& 233 & April 17 & 46 & 0.88 $\times$ 0.73 & 18 & 122$\pm$22\\
7& 343.5 & March 10,26,28 & 120 & 0.66 $\times$ 0.49 & 18 & 67$\pm$19 \\
\hline \noalign {\smallskip}

\end{tabular}
}
\caption{Summary of the ALMA observations of the Vela pulsar. Columns report the observing band, its central frequency $\nu$, the observation date, the integration time, the size of the synthesized beam, rms, and the measured pulsar flux density. We included the error  on the absolute ALMA flux calibration, estimated to be $\sim$10\%, as discussed in Section 2 (see also Fomalont et al.\ 2014).}
\end{center}
\end{table*}

\section{Results}

\begin{figure*}
\centering
{\includegraphics[width=16cm]{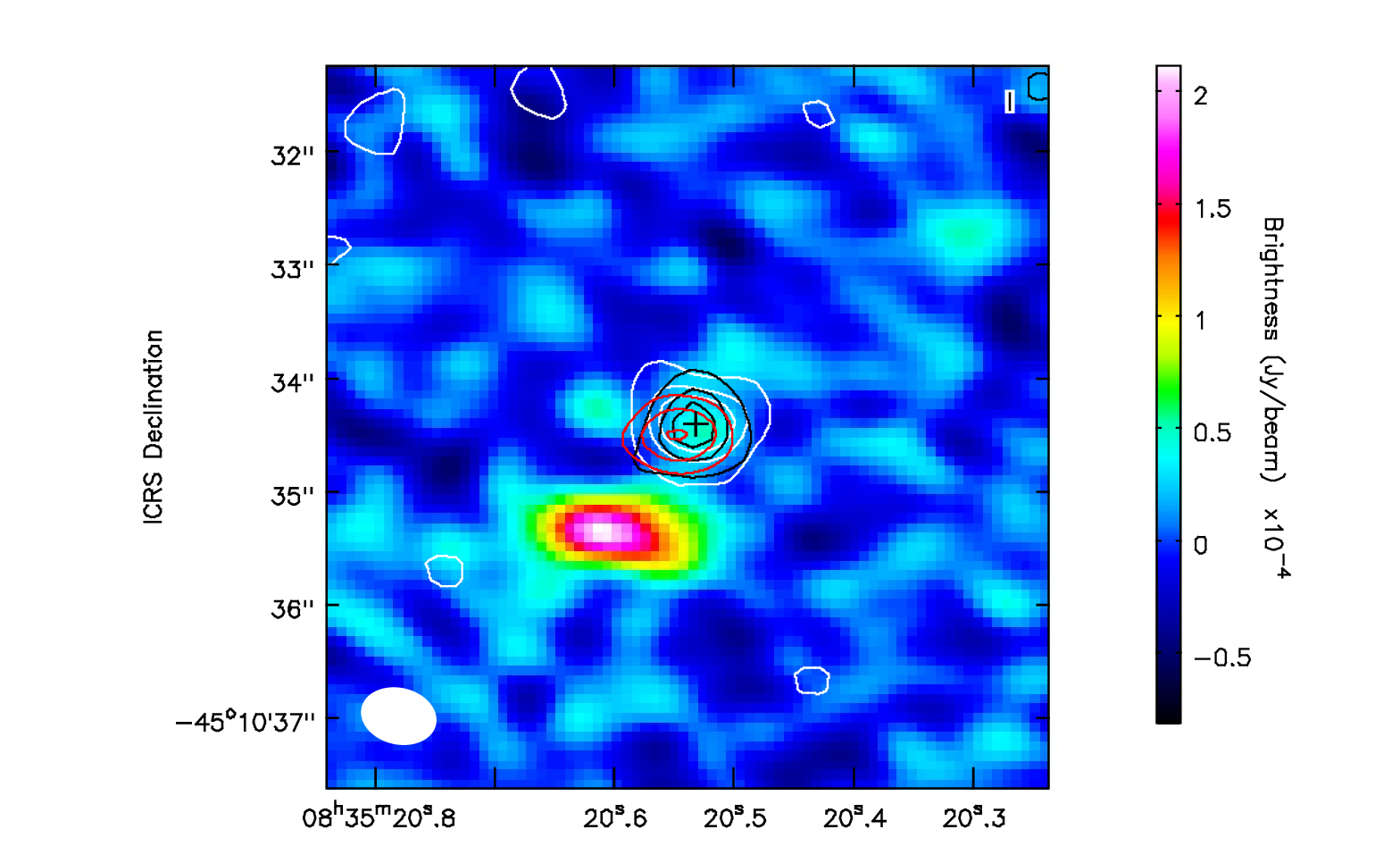}}
\caption{\label{ima} Image of the Vela pulsar field taken with ALMA in Band 7 (343.5 GHz). The  intensity contours overlaid at the centre of the image correspond to the pulsar detections 
in Band 3 (97.5 GHz; white),  Band 4 (145 GHz; black), and Band 6 (233 GHz; red). The contours correspond to 3, 6, 8 $\sigma$ levels for the Band 3 and 4 images, and to 
3, 4, 4.8 $\sigma$ levels for the Band 6 one. The filled ellipse in the left corner represents the size of the synthesised beam of the Band 7 image.
The  extended source about 1\farcs4 southeast of the pulsar is the nebula discussed in Section 4.2. The cross indicates the pulsar position at the average epoch of our ALMA observations.
}
\end{figure*}

The ALMA images in Band 3, 4, and 6  clearly show a source at a position coincident with the radio coordinates of the Vela pulsar,  corrected for its proper motion (Dodson et al.\ 2003).   Figure \ref{ima} shows the ALMA Band 7 image with the source intensity contours from the Band 3, 4, 6 images overlaid, which shows that the source has been detected in all bands.  In all images, its brightness profile is consistent with the instrument point spread function (PSF), as expected from a point source.  The accuracy of the ALMA absolute astrometry makes us confident that we can claim the pulsar identification based upon positional coincidence.
The pulsar is clearly detected, with flux densities of 252$\pm$35, 170$\pm$26, and 122$\pm$22 $\mu$Jy in Band 3, 4, and 6, respectively, including both statistical errors and the uncertainty on the ALMA absolute flux calibration.  It has also been detected in Band 7, although with a lower significance (S/N $\sim$3.7) and a flux density of 67$\pm$19 $\mu$Jy.   We measured the pulsar flux density in the four ALMA  bands within a square aperture of 1\arcsec\ size centred on its position 
 by fitting an elliptical Gaussian to the source brightness profile.  
    
The flux densities of the Vela pulsar in the four ALMA bands 
(Table 1)
can be fitted by a PL with spectral index $\alpha =  -0.93 \pm 0.16$ ($\chi^{2}=0.7$). This clearly indicates that the pulsar emission in the mm/sub-mm range is non-thermal in origin, like in the optical/UV (Romani et al.\ 2005) and in the near-IR (Shibanov et al.\ 2003; Danilenko et al.\ 2011), and is produced in the neutron star magnetosphere.  

An additional, extended and bright source is also visible 1\farcs4 southeast of the pulsar. This source  (hereafter, the "nebula") is only detected in Band 7, where it is relatively bright with a flux of $278\pm 33$ $\mu$Jy computed down to a  $5 \sigma$ intensity contour, where the $\sigma$ is 18 $\mu$Jy/beam (Table 1).

\section{Discussion}

\subsection{The pulsar}

\begin{figure}
{\includegraphics[width=16cm]{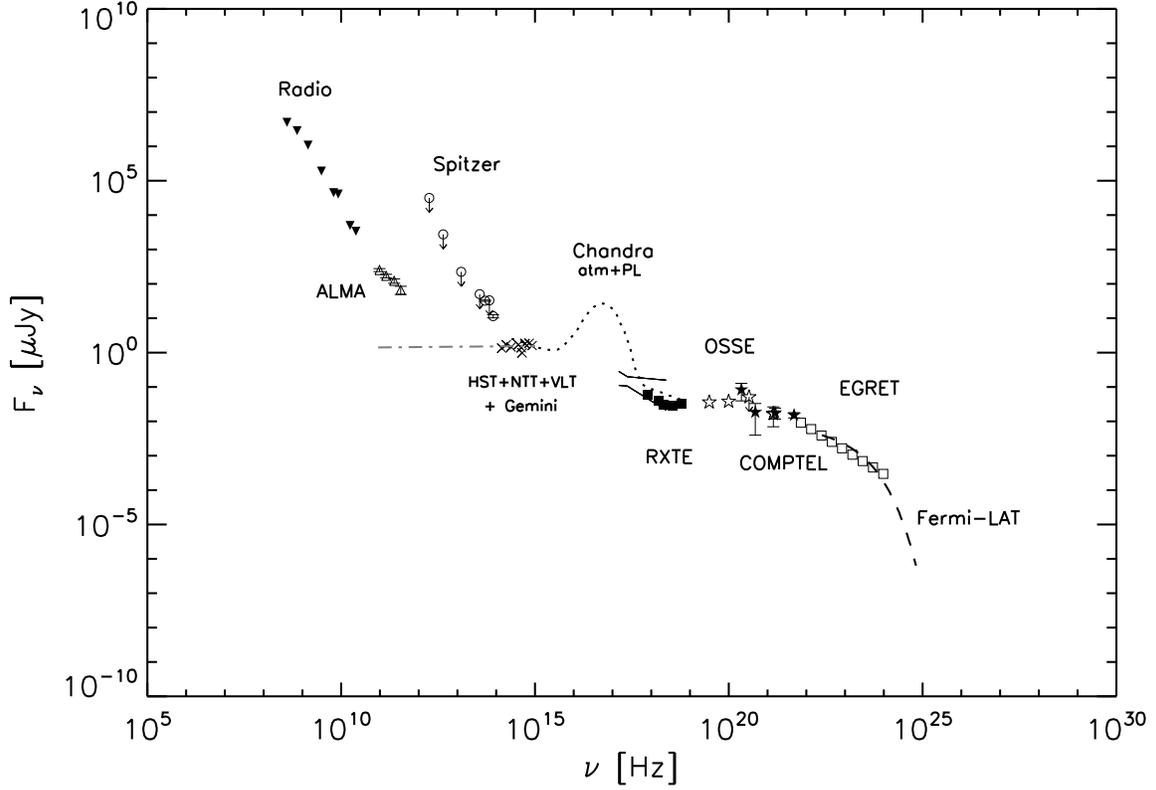}} 
\caption{\label{sed} 
Vela pulsar SED from the radio to the $\gamma$-rays. The radio data (filled inverted triangles)
come from Manchester et al.\ (2005), Johnston et al.\ (2006), Keith et al.\ (2011),  and Johnston (2017, private communication). The open triangles with the error bars are our ALMA data (see Table 1). The rest of the spectrum is built from measurements collected from
the literature (see Danilenko et al.\ 2011; Zyuzin et al.\ 2013 and references therein). The dot-dashed line is the extrapolation of the near-UV--optical--near-IR  PL spectrum (Zyuzin et al. 2013) towards the ALMA bands. The dashed line is the best-fit PL plus exponential cut-off $\gamma$-ray spectrum derived from the {\em Fermi} 
data (Acero et al.\ 2015).
}
\end{figure}

We compared the ALMA  spectrum with the radio spectrum below 24 GHz (Figure \ref{sed}). The latter is presented for eight flux density values in the frequency range between 400 MHz and 24 GHz (Manchester et al.\ 2005; Johnston et al.\ 2006; Keith et al.\ 2011). The spectrum has been recently classified as a broken PL, with the break frequency around 900 MHz, and the spectral index $\alpha = -2.24\pm0.09$ above the break (Jankowski et al.\ 2017).
As shown above, the ALMA spectrum follows a PL with spectral index $\alpha = -0.93\pm0.16$, i.e. it is much flatter than 
the neighbouring radio spectrum. The gap between the two spectra is quite wide, but likely it can be closed smoothly
by e.g., a future detection of Vela at 40 GHz (Band 1) with ALMA.
In the radio domain, the maximal flux density ($\sim$5 Jy) is achieved at $\sim$400 MHz, which yields a brightness temperature  $T_{\rm b} \ga 10^{27}$K for an emitting region smaller than one neutron star radius. Under this assumption,  $T_{\rm b}$ would be 
$\ga 10^{17}$K and $\ga 10^{15}$K, in the ALMA Bands 3 and 7, respectively,  implying coherence.  
Lower, but still quite high, values of $T_{\rm b}$ would be derived if the size of the emission region were, instead, a fraction of the light-cylinder.
A size as large as, e.g.  50 neutron star radii
(Kijak \& Gil 2003) 
would decrease $T_{\rm b}$  by three orders of magnitude. This would not put into question coherent emission at 400 MHz
and in ALMA Band 3 and, possibly,
in Band 7.
Therefore, most likely, coherence  in the mm/sub-mm domain is maintained. 
As a consequence, the transition to incoherent emission processes
must occur at frequencies above 343 GHz, i.e. at wavelengths shorter than 0.87 mm, somewhere between the sub-mm and the near-IR. 
As for the coherent processes responsible for the emission between 400 MHz and 343 GHz, none can be convincingly favoured against another (e.g., Melrose 2017;  Eilek \& Hankins 2016).

We also compared the ALMA PL spectrum of the Vela pulsar with the extrapolation of the flat UV-optical-near-IR PL spectrum (Zyuzin et al.\ 2013) $F_\nu \propto \nu^{0.01 \pm 0.0}$ (Figure \ref{sed}). The latter spectrum also encompasses the flux density (1.35$\pm$0.06 $\mu$Jy) measured in the K$_{\rm S}$-band (138.8 THz) for the pulsar (Zyuzin et al.\ 2013). This flat spectral component is likely due to incoherent magnetospheric synchrotron radiation (SR) by electron-positron pairs. If this is the case, then we can estimate how far the SR component can reach towards lower frequencies. When extrapolated to Band 7, its flux density level would fall short by a factor of $\sim$ 50 below the ALMA point, suggesting  the brightness temperature around $10^{13}$K, i.e. still exceeding the temperature limit of $\sim 10^{12}$K due to the inverse Compton catastrophe (Kellerman \& Pauliny-Toth 1969). We conclude, therefore, that  the UV--optical--near-IR spectrum due to incoherent SR does not extend to the sub-mm range but it should turn over somewhere between the mid- and far-IR due to either synchrotron self-absorption or low-energy cutoff.
Finally, we compared the ALMA PL spectrum of the Vela pulsar with the 
extrapolation of the 
mid-IR flux densities (Danilenko et al.\ 2011) to the mm/sub-mm region\footnote{The {\em Spitzer}  mid-IR flux measurements at 3.6 and 5.8 $\mu$m  (83275 and 51688 GHz, respectively) are uncertain owing to the possible contamination of a structure at 1\farcs4, which can not be spatially resolved from the pulsar at the  {\em Spitzer} angular resolution (Zyuzin et al.\ 2013). }. The ALMA spectrum is clearly below such an extrapolation (Figure \ref{sed}), which indicates a turnover in the  wavelength interval between $\sim 0.87$ mm and 5.8 $\mu$m 
(343 and 51688 GHz, respectively). 

Changes in the PL slope from the UV to the radio are also seen in the Crab pulsar (e.g., Danilenko et al.\ 2011). Whether these changes reflect emission from different population of relativistic particles, a difference in the location of the emission regions in the magnetosphere,  or different emission mechanisms
 can only be clarified by further studies.

We also note that the slope of the ALMA spectrum rules out a possibility that  the  mid-IR--to--sub-mm spectrum of the Vela pulsar is described by a low-temperature blackbody, with the ALMA fluxes along its Rayleigh-Jeans tail, which could be associated with emission from a cold debris disc  left over from the supernova explosion  (\"Ozs\"ukan et al.\ 2014). The non detection of a disc around the Vela pulsar may support the hypothesis that high inclination angles $\alpha$ between the pulsar magnetic dipole and spin axis yield unfavourable conditions for such discs to be stable. A compilation of the values of $\alpha$, based on the combination of both radio and $\gamma$-ray observations, is given in Barnard et al.\ (2016), and shows that they are mostly close to $70^{\circ}$. 
A value of $\alpha \sim 53^{\circ}$ is inferred by Johnston et al.\ (2005) but using radio observations only.
 
\subsection{The Nebula}

\begin{figure*}
\centering
{\includegraphics[width=8cm]{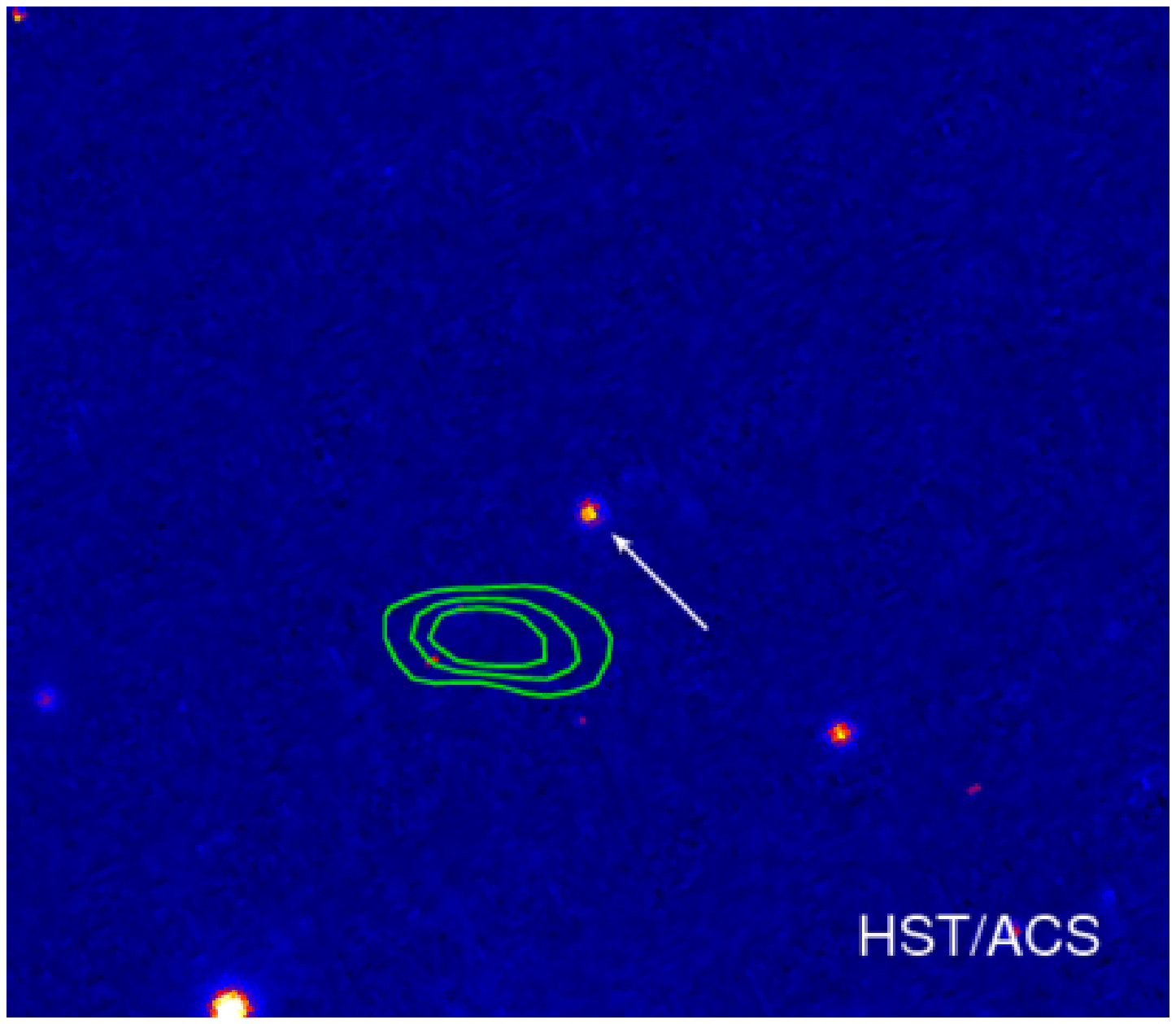}} 
{\includegraphics[width=8cm]{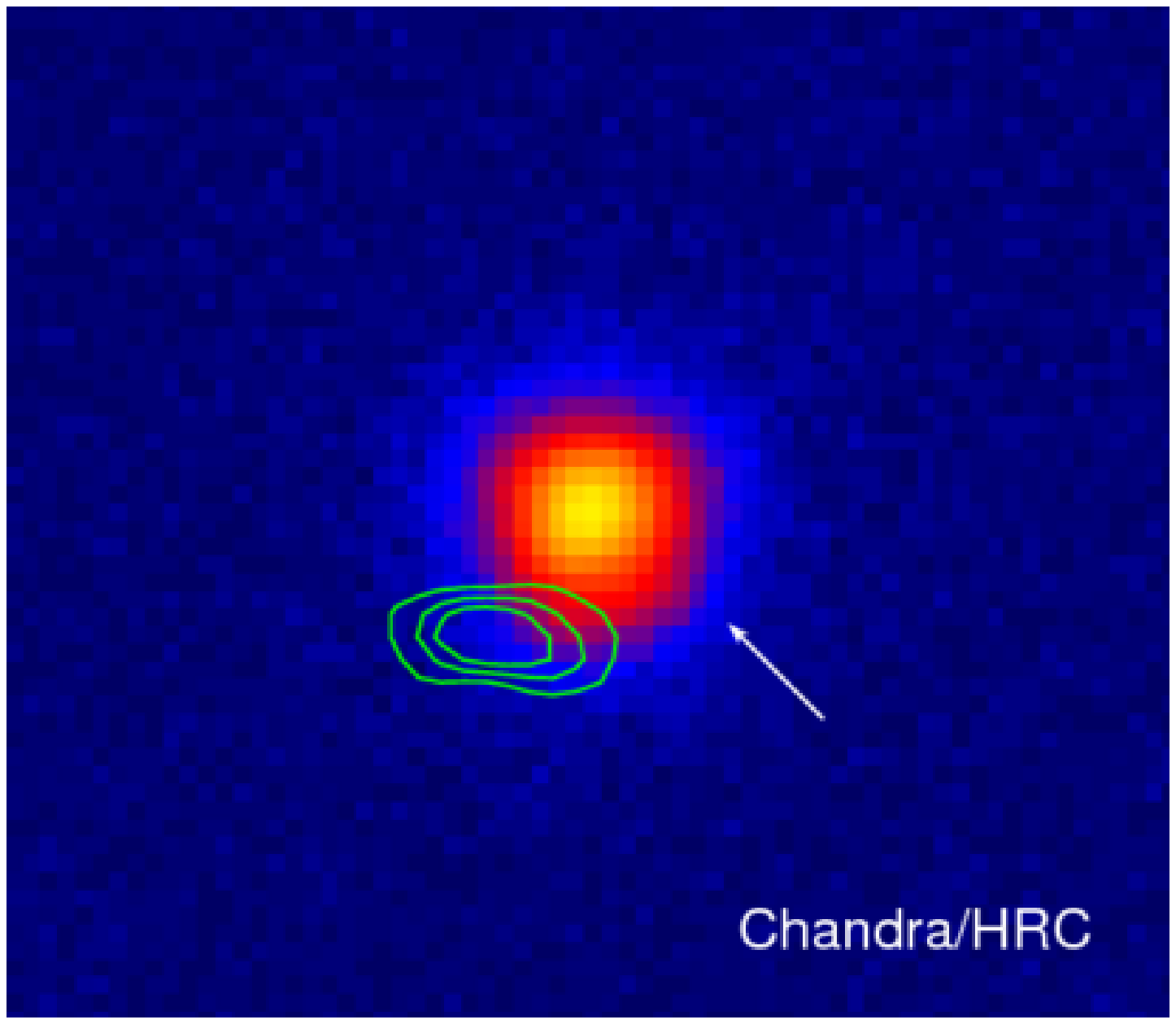}} 
{\includegraphics[width=8cm]{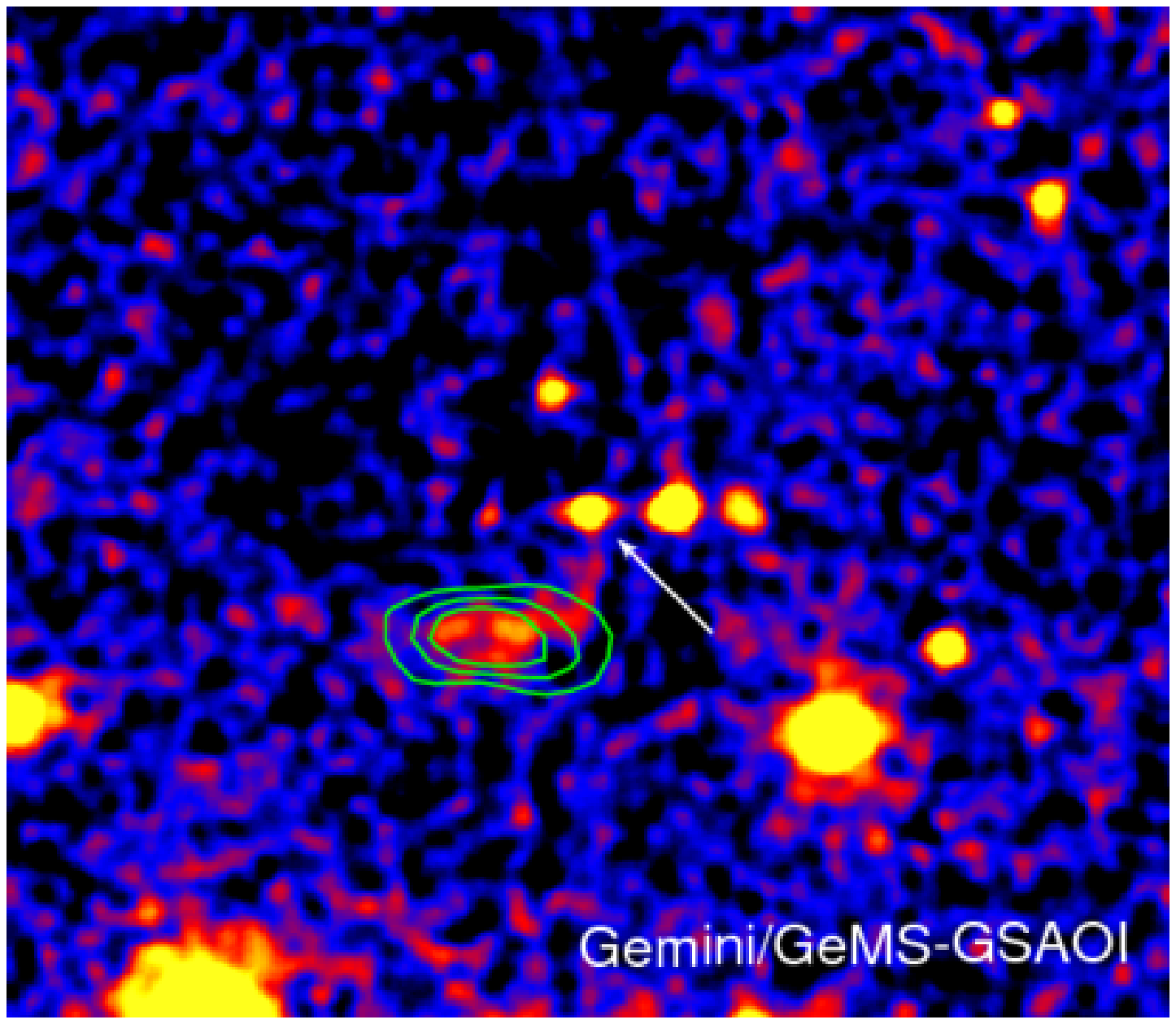}} 
{\includegraphics[width=8cm]{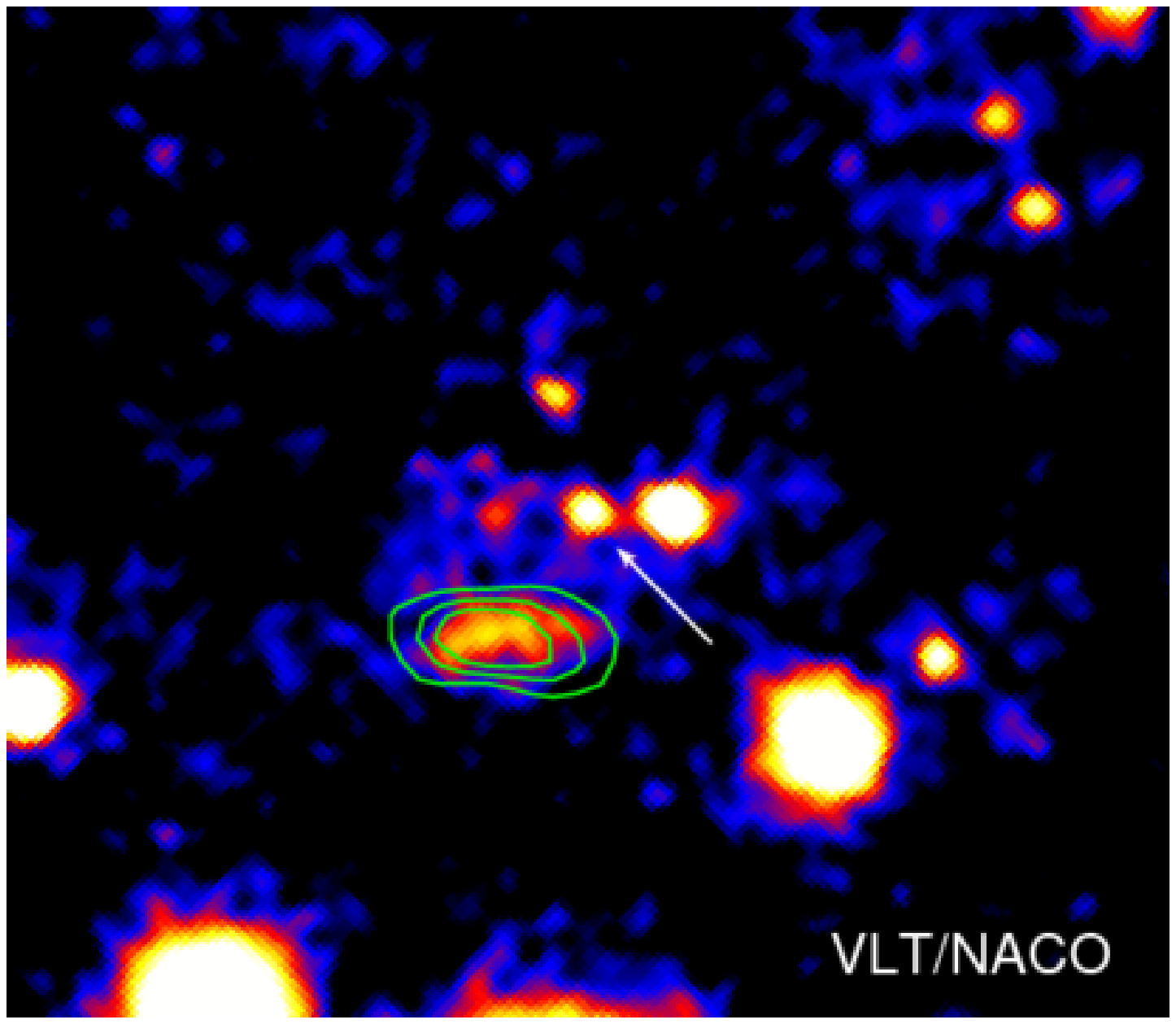}} 
\caption{\label{blob}   ALMA Band 7 intensity contours (in green) of the nebula close to the Vela pulsar (see Figure \ref{ima}) superimposed on the {\em HST}/ACS, {\em Chandra}/HRC, Gemini/GeMS-GSAOI, and VLT/NACO images of the field, top left to bottom right. In all panels the angular dimension is 10\arcsec$\times$10\arcsec\ and  the arrow indicates the pulsar. The time baseline between the Gemini, VLT, and ALMA observations ($\sim$ 3 years), together with the spatial resolution (0\farcs8) and astrometry accuracy (0\farcs123) of the Band 7 image, 
are not sufficient to appreciate a relative angular displacement of the nebula comparable to that predicted for the pulsar ($\approx$ 0\farcs2) as a results of its proper motion.}

\end{figure*}

We compared the ALMA Band 7 image with 
those taken with the Advanced Camera for Surveys (ACS) aboard {\em HST} and the 606W filter ($\lambda=0.59 \mu$m; $\Delta \lambda=0.25\mu$m), see Figure \ref{blob} (top left).
There is no stellar object close to the position of the nebula. Therefore, it is unlikely that it is associated with the unresolved contribution of background stars.  
We note that the position of the nebula emission centroid with respect to the pulsar position lies close to the axis of symmetry of the X-ray PWN observed by {\em Chandra}, which is at a position angle of $310^{\circ} \pm 1.5^{\circ}$, computed east from north (Helfand et al.\ 2001).  To investigate a possible association with the PWN we 
compared the ALMA Band 7 image
with a {\em Chandra} one  (Figure \ref{blob}, top right)
taken with the High Resolution Camera in imaging mode (HRC-I;
Obs. ID 1966).
However, even exploiting the sharp {\em Chandra} PSF, the nebula is too close to the pulsar to resolve possible X-ray features on an angular scale as small as 1\farcs4. 

Interestingly,  
based on Gemini-South K$_{\rm s}$-band images taken in January 2013 with the Gemini Multi-Conjugate Adaptive Optics System (GeMS) and the Gemini South Adaptive Optics Imager (GSAOI)
Zyuzin et al.\ (2013) claimed the detection of an extended structure, also at $\sim 1\farcs4$ south-east from Vela and aligned closely with its PWN symmetry axis, which they interpreted as a counter-jet protruding from the pulsar.  
To confirm the Gemini detection, we used archival\footnote{www.eso.org/archive}, unpublished, 
K$_{\rm s}$-band high spatial resolution  images of the Vela pulsar field taken between December 2012 and February 2013 with the S54 camera (54\arcsec$\times$54\arcsec; 0\farcs054 pixel scale)  of the Nasmyth Adaptive Optics System (NAOS) Near-IR Imager and Spectrograph (CONICA) -- NACO -- at the VLT.  Figure \ref{blob} (bottom left) shows the Gemini/GeMS-GSAOI K$_{\rm s}$-band image compared with the almost contemporary VLT/NACO K$_{\rm s}$-band one (bottom right). 
The same structure seen in the former image is also detected at exactly the same position in the latter,  confirming that it is real and not due to a background fluctuation.  

The nebula seen in the ALMA Band 7 image is similar
in morphology to the structure detected in the  K$_{\rm s}$-band images and is spatially coincident with it (Figure \ref{blob}, bottom), meaning that they are the same object.  
One possibility is that it would form 
from a jet-like activity from the pulsar, perhaps a counter-jet (Zyuzin et al.\ 2013) like that observed in the X-ray PWN.
At the distance of the Vela pulsar of  $287^{+19}_{-17}$ pc (Dodson et al.\ 2003), the Band 7 luminosity of the nebula would be  $\sim 6 \times 10^{25}$ erg s$^{-1}$, 
which is a tiny fraction of the pulsar rotational energy loss rate $\dot{E} = 6.96\times10^{36}$ erg s$^{-1}$.
An emission knot like that observed close to the Crab pulsar (Hester et al.\ 1995)  might be another possibility.
Very little can be said on the nebula spectrum, with only a direct K$_{\rm s}$-band detection (2.34$\pm$0.17; $\mu$Jy; Zyuzin et al.\ 2013) to be compared with our Band 7 one ($278\pm 33$ $\mu$Jy), which corresponds to a PL with $\alpha = -0.79 \pm 0.02$.

\section{Summary and Conclusions}

We detected the Vela pulsar with ALMA at frequencies of 97.5, 145, 233, and 343.5 GHz. This is the first detection of a pulsar ever obtained with ALMA. We found that the pulsar spectrum  is fitted by a PL with  spectral index $\alpha  =  -0.93 \pm 0.16$. Our observations show that the brightness temperature is still quite high with values of $ \approx 10^{27}$K  at 300 MHz down to $10^{15}$ K at 343 GHz (0.87 mm), for an emission region of the same size as the pulsar,
indicating that coherence  is maintained.
Finding pulsations at the radio period, or placing strong constraints on their presence, in one or more of the ALMA bands by exploiting the phased-array mode  would be a strong test for a connection or lack thereof  between the mm/sub-mm emission and the radio emission from the comparison of the pulse profiles.  If the ALMA flux is pulsed, with a pulse profile similar to the radio one, this would imply that the radio emission still extends to the mm/sub-mm region. At the same time, detecting pulsations would be a firm identification evidence.

Our results show that  the transition to incoherent emission processes, which are expected to dominate at least from the near-IR up to the highest energies, must occur at wavelengths shorter than 0.87 mm, somewhere between the sub-mm and the near-IR.   
Measuring the mid-IR spectrum of the Vela pulsar between $\sim$0.005 mm and $\sim$0.03 mm with the  {\em James Webb Space Telescope (JWST)} would then be crucial to constrain the wavelength range where such a transition occurs. 

Our ALMA observations, together with archival VLT ones in the near-IR, also confirmed the existence of an extended emission structure (the "nebula") at 1\farcs4 from the pulsar (Zyuzin et al.\ 2013). 
 {\em JWST} observations would better characterise its spectrum in the near-IR/sub-mm.
Monitoring in time the position, brightness, and morphology of the nebula would also provide  information on its origin and its connection with the pulsar.

\acknowledgments

We dedicate this work to our dear friend Giovanni F. Bignami, who passed away unexpectedly when this manuscript was in preparation. Nanni, as he was known to his friends, has played a major role in promoting multi-wavelength observations of pulsars and  authored many inspirational works on the study of the Vela pulsar. This paper makes use of the following ALMA data: ADS/JAO.ALMA\#2015.1.01104.S. ALMA is a partnership of ESO (representing its member states), NSF (USA) and NINS (Japan), together with NRC (Canada), MOST and ASIAA (Taiwan), and KASI (Republic of Korea), in cooperation with the Republic of Chile. The Joint ALMA Observatory is operated by ESO, AUI/NRAO and NAOJ. Based on observations made with ESO Telescopes at the La Silla Paranal Observatory under programme ID 090.D-0425(A). RPM acknowledges financial support from an "Occhialini Fellowship". BR acknowledges financial support by the National Science Centre
Grant DEC-2011/02/A/ST9/00256. We thank Simon Johnston for providing information on the radio flux densities of Vela and Patrizia Caraveo for her careful reading of the manuscript.

{\it Facilities:} \facility{Atacama Large Millimetre Array, Very Large Telescope}

\end{document}